\documentclass[runningheads]{llncs}
\usepackage[T1]{fontenc}
\usepackage{graphicx,verbatim,subcaption,amsmath,xcolor}
\begin{document}
\title{Weighted Mean Frequencies: a handcraft Fourier feature for 4D Flow MRI segmentation}
\titlerunning{WMF: a handcraft Fourier feature for 4D Flow MRI segmentation}
%
%\begin{comment}  %% Removed for anonymized MICCAI 2025 submission
\author{Simon Perrin\inst{1}\orcidID{0009-0005-2008-2802} \and
Sébastien Levilly\inst{1}\orcidID{0000-0002-2704-0315} \and
Huajun Sun\inst{2} \and %TODO orcid de Huajun
Harold Mouchère\inst{1}\orcidID{0000-0001-6220-7216} \and
Jean-Michel Serfaty\inst{3}\orcidID{0000-0001-8801-1414}
}
\authorrunning{S. Perrin et al.}
% First names are abbreviated in the running head.
% If there are more than two authors, 'et al.' is used.
%
\institute{Nantes Université, École Centrale Nantes, CNRS, LS2N, UMR 6004, F-44000 Nantes, France\\
\email{\{simon.perrin,sebastien.levilly,harold.mouchere\}@univ-nantes.fr} \and
Nantes Université, CHU Nantes, INSERM, CNRS, CRCI2NA, F-44000 Nantes, France\\
\email{huajun.sun@chu-nantes.fr}\and
CHU Nantes, CNRS, INSERM, L'Institut du Thorax, Nantes Université, 44000 Nantes, France\\
\email{jeanmichel.serfaty@chu-nantes.fr}}

%\end{comment}

%\author{Anonymized Authors}  %% Added for anonymized MICCAI 2025 submission
%\authorrunning{Anonymized Author et al.}
%\institute{Anonymized Affiliations \\
%    \email{email@anonymized.com}}

\def\x{{\mathbf x}}
\def\L{{\cal L}}
\def\4d{{4D Flow MRI}}
\def\FT{\text{FT}}
\def\WMF{\text{WMF}}

\newcommand{\seb}[1]{\textcolor{red}{#1}}
\newcommand{\sebl}[1]{\textcolor{magenta}{#1}}
\newcommand{\sebc}[1]{\textcolor{brown}{#1}}
\newcommand{\harold}[1]{\textcolor{blue}{#1}}
\newcommand{\haroldl}[1]{\textcolor{cyan}{#1}}
\newcommand{\simon}[1]{\textcolor{violet}{#1}}
\newcommand{\simonl}[1]{\textcolor{teal}{#1}}

\maketitle              % typeset the header of the contribution
\begin{abstract} % 150-250  words

In recent decades, the use of \4d images has enabled the quantification of velocity fields within a volume of interest and along the cardiac cycle.
However, the lack of resolution and the presence of noise in these biomarkers are significant issues.
As indicated by recent studies, it appears that biomarkers such as wall shear stress are particularly impacted by the poor resolution of vessel segmentation.
The Phase Contrast Magnetic Resonance Angiography (PC-MRA) is the state-of-the-art method to facilitate segmentation.
The objective of this work is to introduce a new handcraft feature that provides a novel visualisation of \4d images, which is useful in the segmentation task.
This feature, termed Weighted Mean Frequencies (WMF), is capable of revealing the region in three dimensions where a voxel has been passed by pulsatile flow.
Indeed, this feature is representative of the hull of all pulsatile velocity voxels.
%
%The interest of that feature is demonstrated by two experimentations to segment \4d images with a manual thresholding and deep-learning method.
The value of the feature under discussion is illustrated by two experiments. 
%
%The first experiment involved the segmentation of 4D images using a manual \harold{optimal} thresholding method. 
%
%The second experiment involved the segmentation of 4D images using a deep learning method.
%
The experiments involved segmenting \4d images using optimal thresholding and deep learning methods.
The results obtained demonstrate a substantial enhancement in terms of IoU and Dice, with a respective increase of 0.12 and 0.13 in comparison with the PC-MRA feature, as evidenced by the deep learning task.
This feature has the potential to yield valuable insights that could inform future segmentation processes in other vascular regions, such as the heart or the brain.

%The abstract should briefly summarize the contents of the paper in 150--250 words.  If you are to include a link to your Repository, please make sure it is anonymized for the double-blind review phase.

\keywords{4D Flow MRI \and Handcraft Feature \and Segmentation.}
% Authors must provide keywords and are not allowed to remove this Keyword section.

\end{abstract}

\section{Introduction}
\label{sec:intro}

\4d is a promising sequence that provides the anatomy and velocity field in a 3D volume and along the cardiac cycle~\cite{Markl2012}.
The quantification of hemodynamic biomarkers, such as wall shear stress (WSS), is enabled by these 3D measurements~\cite{Levilly2020}.
%These 3D measurements allow the quantification of hemodynamic biomarkers such as the wall shear stress~\cite{levillyQuantitativeEvaluationWall2020}.
%
%Unfortunately, such parietal biomarkers necessitates the precise vessel segmentation which is a tedious task for clinician.
The necessity of vessel segmentation for parietal biomarkers, such as WSS, is a source of considerable time loss for clinicians.
%
%However, state-of-the-art deep-learning approach can tackle successfully this task on specific vessel or organs and at the price of a large annotated dataset~\cite{Markl1000aorta,BustamanteV2,PerezGarcia}.
Nevertheless, state-of-the-art deep-learning approaches have been successfully employed to address this task for specific vessels or organs, necessitating a substantial annotated dataset~\cite{berhaneFullyAutomated3D2020,Bustamante2023,MARINCASTRILLON202320}.

Clinicians and state-of-the-art solutions frequently use a handcrafted feature, designated Phase-Contrast Magnetic Resonance Angiography (PC-MRA), which facilitates the distinction between flows, vessels and cavities~\cite{BUSTAMANTE2018128,dumoulinPCMRA}.
%Clinician and state-of-the-art solutions~\cite{BustamanteV1,PerezGarcia,algo qui prends PC-MRA en entrée} often rely on a handcraft feature, named Phase-Contrast Magnetic Resonance Angiography (PC-MRA), that facilitates the distinction between flows, vessels and cavities~\cite{Dumoulin}.
%
%PC-MRA images is function of the anatomical image and the norm of the velocity field.
The PC-MRA image is the result of the anatomical image and the norm of the velocity field multiplication.
%
%It can be computed for each time frame but its use is often dedicated to the systolic time.
Its calculation is possible for each time frame; however, its application is frequently focused on the systolic time.
Indeed, diastolic PC-MRA does not benefit from high velocity field and is then degraded by noise.
Furthermore, the absence of contrast between the lumen and the background can have a substantial impact on the PC-MRA image~\cite{pcmrabock2010}.
%\simon{Ajouter un petit mot sur le fait que la qualité de PC-MRA est aussi lié au contraste de l'image anatomique et donc, de la présence d'un agent de contraste} 
%
Segmentation of the diastolic period is consequently a challenging issue.

The objective of this study is to propose a new handcraft feature, Weighted Mean Frequencies (WMF), that assists with the segmentation task, whether performed manually or automatically.
%This study aims to propose a new handcraft feature, named Weighted Mean Frequencies (WMF), that helps with the segmentation task, either manually or automatically.
%
As outlined in Sec.~\ref{sec:wmf}, the purpose of this feature is to depict voxel-wise the predominant blood flow frequencies in a 3D volume.
%This feature objective, described in Sec.~\ref{sec:wmf}, is to represent in a 3D volume the most contributing blood flow frequencies by voxel.
%
Consequently, WMF is capable of discerning voxels that have experienced pulsatile flow during a specific phase of the cardiac cycle.
%Therefore, WMF illustrates voxels that observed a pulsatile flow during a period of the cardiac cycle.
%
In order to demonstrate the practical benefits of WMF, a comprehensive investigation into two segmentation tasks is presented in Sec.~\ref{sec:manual_seg} and~\ref{sec:deep_seg}. The first of these is a threshold segmentation task, and the second is a deep-learning-based segmentation task.
%To demonstrate the utility of WMF, a detailed study of two segmentation tasks is presented in Sec.~\ref{sec:manual_seg} and~\ref{sec:deep_seg} : a threshold segmentation task and deep-learning based segmentation task.
%
%Finally, these results are discussed with regards of the PC-MRA images.
In conclusion, the results are discussed in relation to the PC-MRA images.

%\seb{Essor de la 4D Flow et intérêt clinique (biomarqeurs)}

%\seb{Intérêt de la segmentation pour ces biomarqueurs.}

%\seb{Complexité pour la segmentation par des cliniciens (temps,tâche ingratte), difficulté pour des outils automatique de distinguer la séparation entre différents vaisseaux en raison de la résolution. Acquisition longue avec des difficultés/variabilités de gestion de l'injection de contraste.}

%\seb{Les cliniciens et algo s'appuient sur PC-MRA (ref) pour sortir une segmentation => résultat intéressant et performant sauf dans le cas des vitesses plus faibles ou principalement validés sur des temps systoliques (regarder publi Bustamante)}

%\seb{PC-MRA par définition est influencé par le signal anatomique (et son contraste) et la quantitée de vitesse au temps donné.}

%\seb{Possibilité d'utiliser le PC-MRA systolique pour d'autres temps mais au risque d'avoir des erreurs significatives sur les zones particulièrement mobiles.}

%\seb{Notre contribution : volonté de caractériser les zones ayant vu un écoulement pulsée pour construire une enveloppe des voxels d'intérêts (facilitant la segmentation)}

%\seb{Présenter la méthodologie démontrant l'intérêt de la feature}

%\seb{Ouverture : Ne s'oppose pas nécessairement à PC-MRA ?}

%\section{Related works}
%\label{sec:related_works}

\section{Weighted Mean Frequencies}
\label{sec:wmf}

Weighted Mean Frequencies (WMF) are computed as the mean of the energy-weighted frequencies of the temporal Fourier Transform (FT).
Firstly, the FT energies vector $\text{E}(u_j)$ is defined as, $|\FT(u_j)|^2$, %the square of the absolute FT values of a speed component :
%\begin{equation}
%    $E_{\FT}(u) = |\FT(u)|^2$
%\end{equation}
where $u_j$ is a velocity component at the $j$-th voxel position.
%speed image
%and the FT is performed along the temporal axis.
%
Moreover, WMF is calculated by averaging strictly positive frequencies weighted by their energies, as
%mean is performed on the strictly positive FT frequencies weighted by the energy, as:
\begin{equation}
    \WMF(u_j) = \frac{\sum^n_{i=1}\text{E}_{i}(u_j)  f_i}{\sum^n_{i=1}\text{E}_{i}(u_j)},
\end{equation}
where $f_i$ is the $i$-th frequency and $n$ is the number of strictly positive frequencies.
%
%As we are computing WMF for each component, we want to merge them together to have one feature that represent the overall fluid energy.

In the following, WMF will represent the 3D volume of the most contributing frequency at voxel level.
%
%As WMF can be computed for each component, the objective is to merge them together to create a single feature that represents the overall fluid pulsatility energy.
Since WMF can be computed per component, the goal is to merge them into a single feature representing overall fluid pulsatility energy.
%
%As merging strategy, we chose to use the $Min$ function to save the minimum frequencies that are present in the fluid domain:
%It is important to note that a high WMF frequency is indicative of a greater noise contribution. 
A high WMF frequency is observed to reflect increased noise contribution.
%A high WMF indicates a greater noise contribution.
%
Therefore, the minimum operator is applied over the WMF for each component:
\begin{equation}
    \WMF_\text{min} = \text{min}(\WMF_\text{u}, \WMF_\text{v}, \WMF_\text{w})\label{eq:WMFmin},
\end{equation}
where $\WMF_\text{min}$ is the minimum WMF of the velocity components (u, v, w). 

\begin{figure}[tbh!]
    \centering
    \begin{subfigure}{0.192\linewidth}
    \centering
        \includegraphics[width=\textwidth]{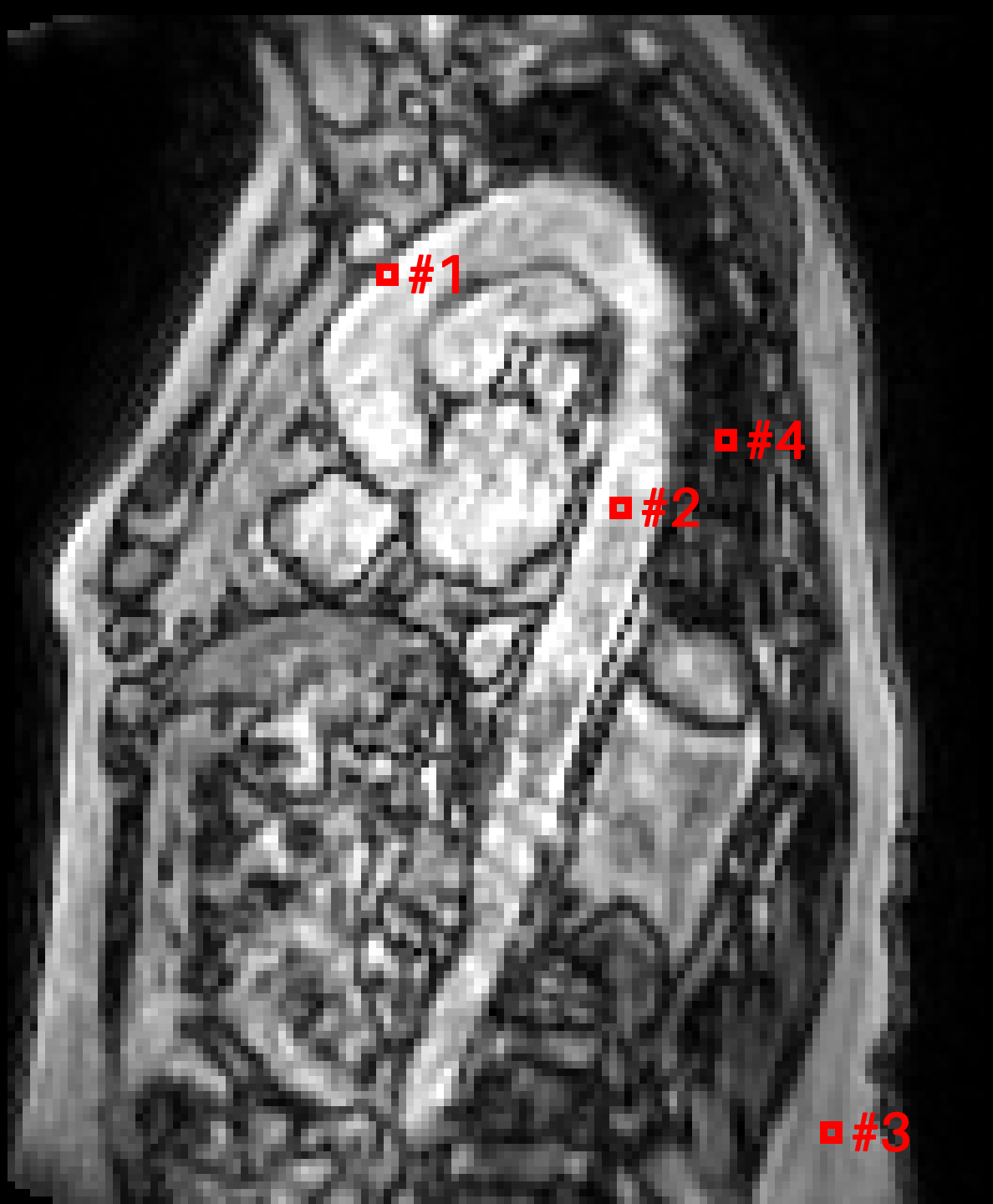}%
        \caption{Anatomy}
        \label{fig:analysis_wmf_a}
    \end{subfigure}
    \begin{subfigure}{0.192\linewidth}
        \includegraphics[width=\textwidth]{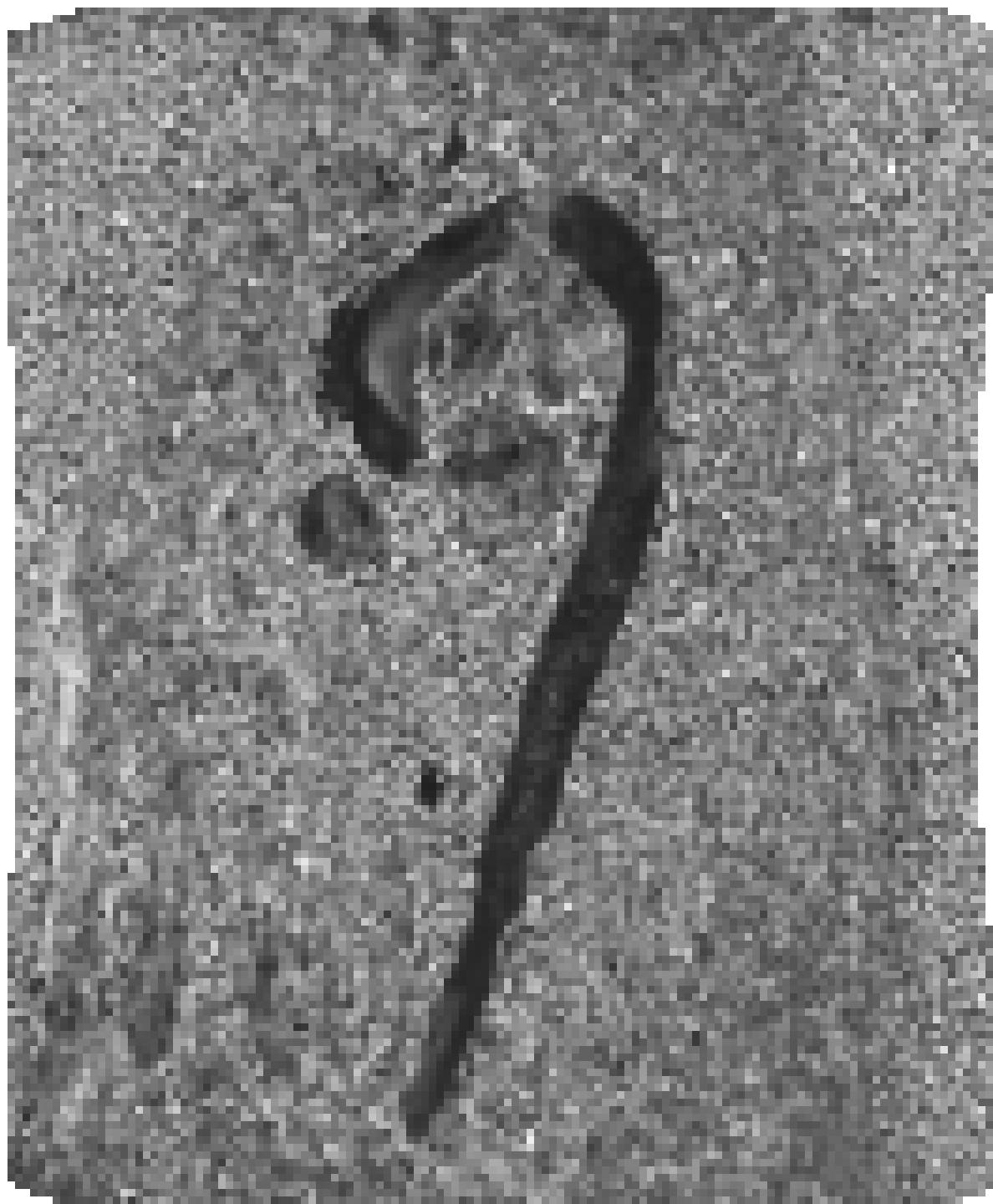}%
        \caption{$\WMF_\text{u}$}
        \label{fig:WMF_a}
    \end{subfigure}
    \begin{subfigure}{0.192\linewidth}
        \includegraphics[width=\textwidth]{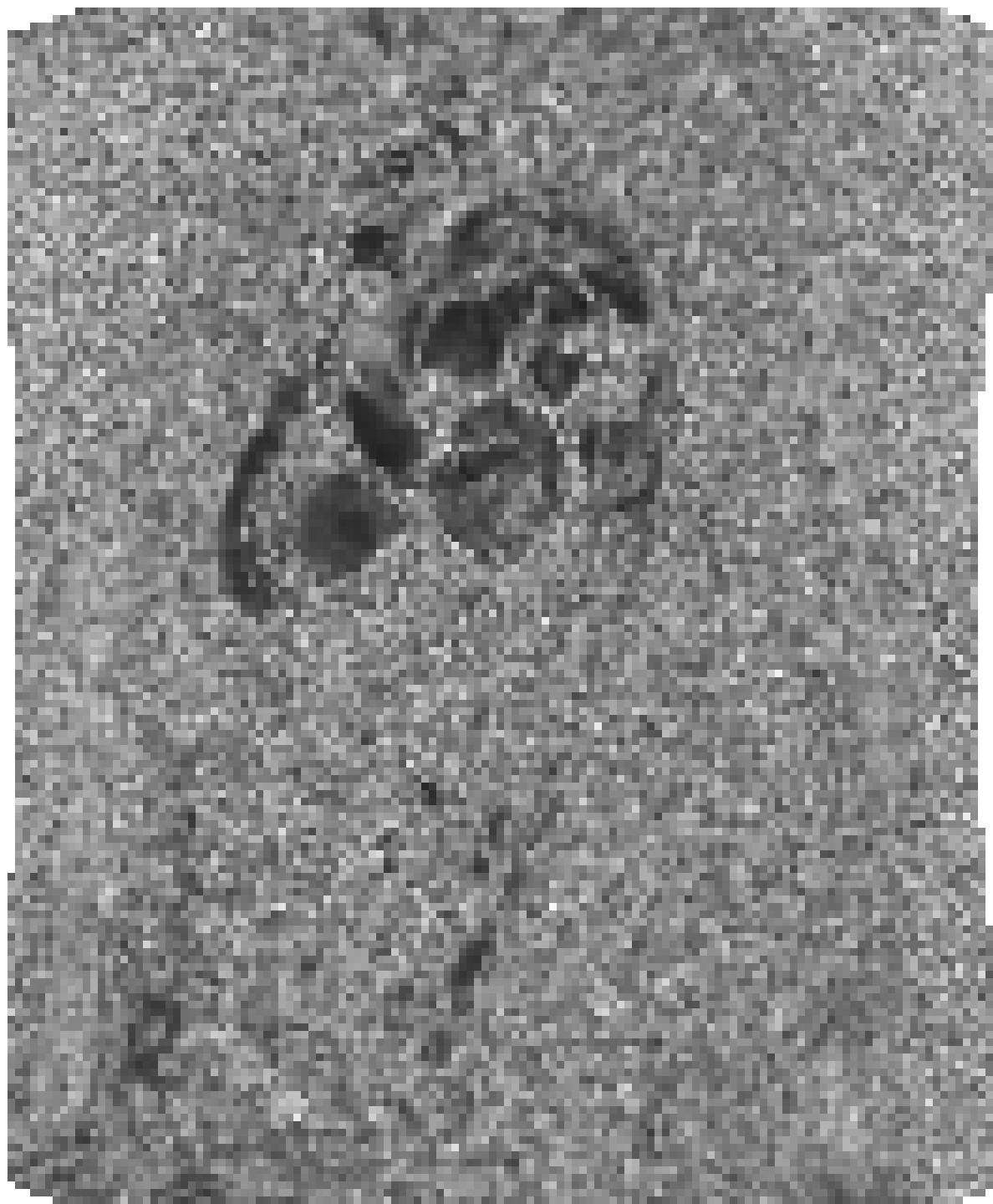}%
        \caption{$\WMF_\text{v}$}
        \label{fig:WMF_b}
    \end{subfigure}
    \begin{subfigure}{0.192\linewidth}
        \includegraphics[width=\textwidth]{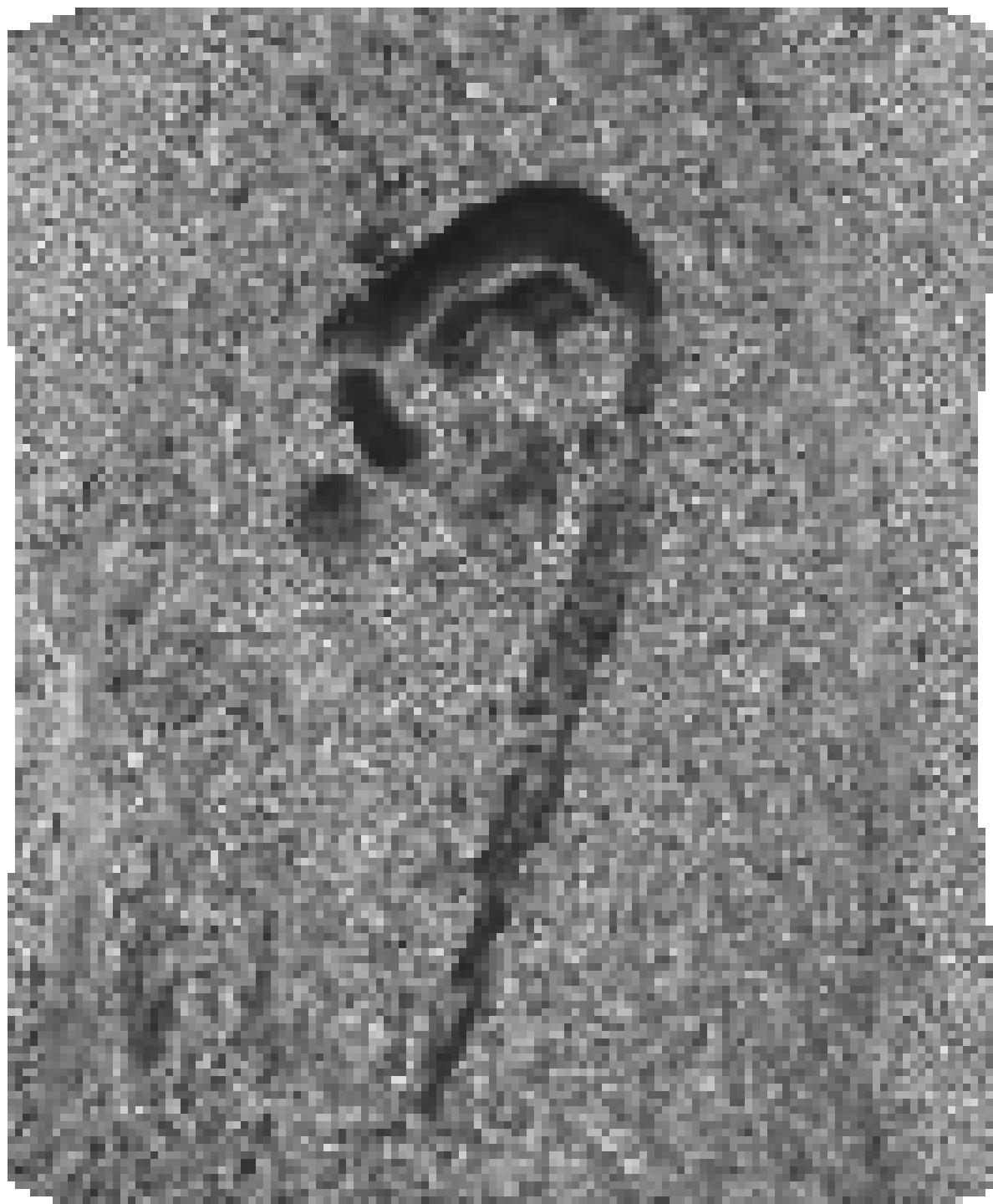}%
        \caption{$\WMF_\text{w}$}
        \label{fig:WMF_c}
    \end{subfigure}
    \begin{subfigure}{0.192\linewidth}
        \includegraphics[width=\textwidth]{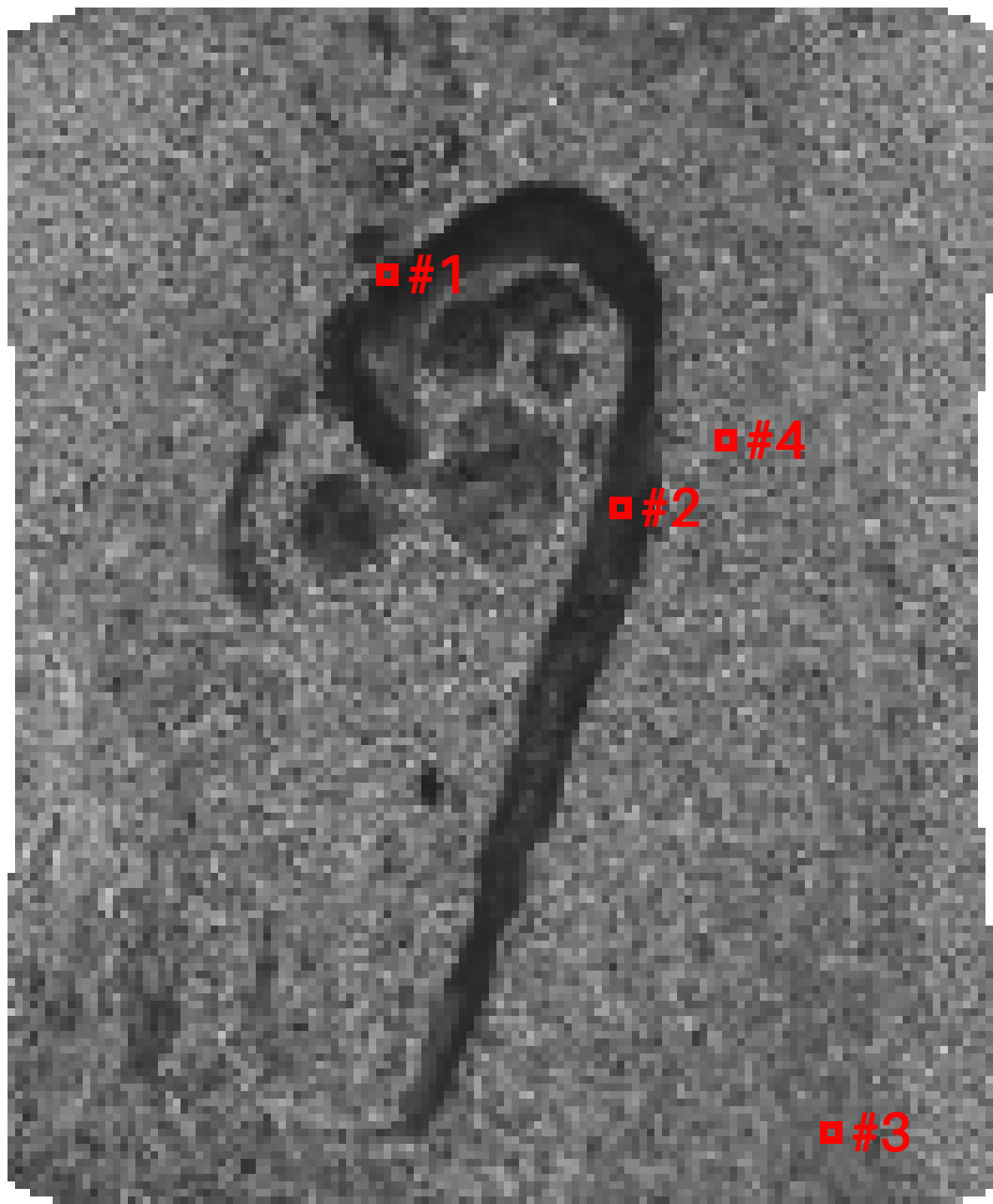}%
        \caption{$\WMF_\text{min}$}
        \label{fig:analysis_wmf_b}
    \end{subfigure}\\
    \begin{subfigure}{\linewidth}
        \centering\includegraphics[width=\textwidth]{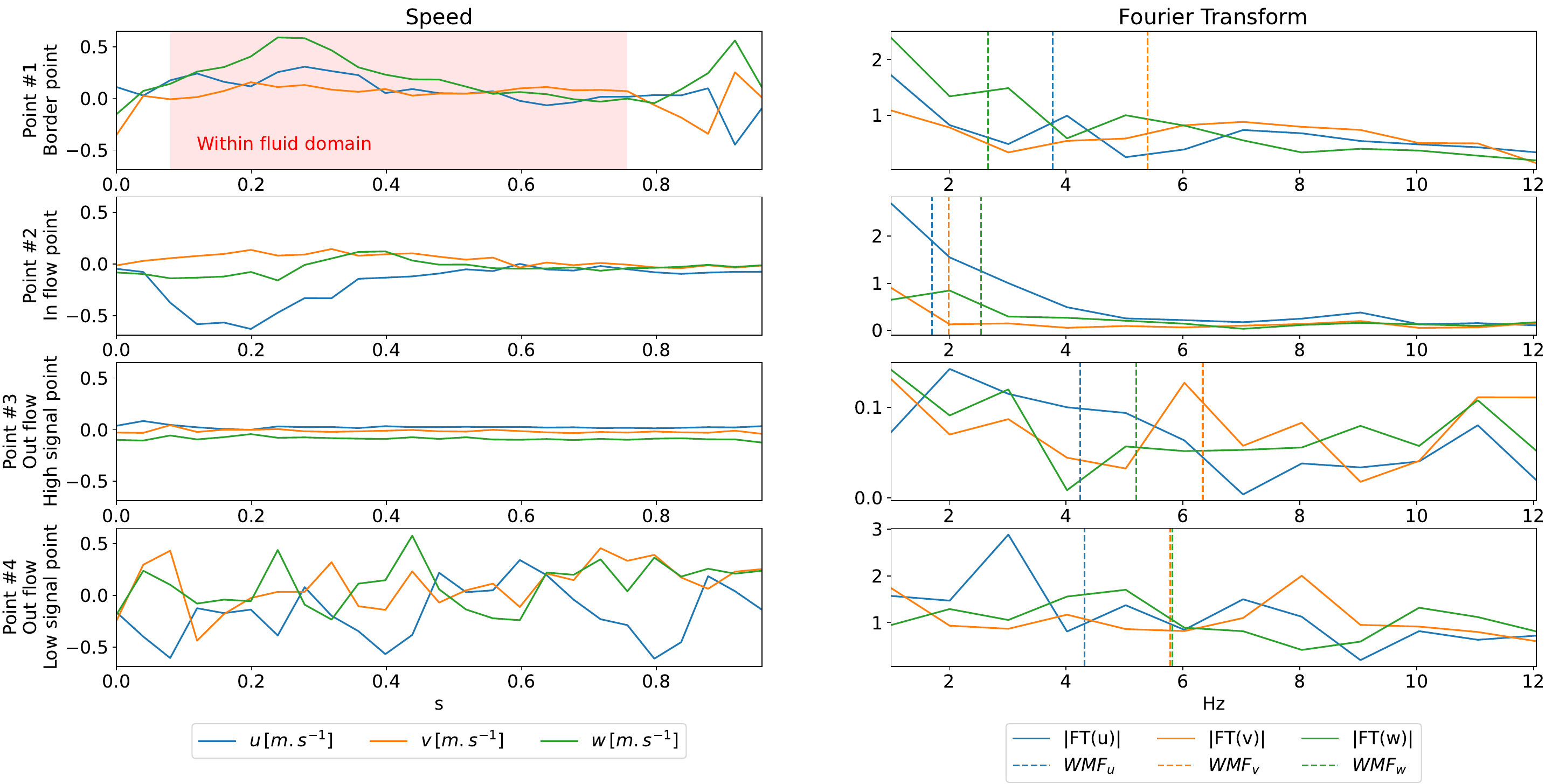}%
        \caption{Speed and Fourier Transform of each point (strictly positive frequencies)}
    \label{fig:analysis_wmf_c}
    \end{subfigure}
    \caption{Images (a) shows the anatomy, (b), (c) and (d) represent the result of WMF for each phase component, (e) is the image obtained using $\WMF_\text{min}$ and (f) is speed and Fourier Transform analysis at different point in a \4d.}
    \label{fig:WMF}
\end{figure}

In order to illustrate the behaviour of the proposed feature on \4d, Figure~\ref{fig:WMF} presents the anatomy, WMF of each component, $\WMF_\text{min}$ and observation graphs. 
%
%The latter describes the WMF behaviour on different point positions.
%
Figure~\ref{fig:analysis_wmf_a} shows the anatomy of an patient with 4 points located on different region of interest.
%
%Figure~\ref{fig:WMF} presents the result of applying WMF on a \4d, where first three images are images obtained for WMF on each velocity component and the last on is about the $\WMF_\text{min}$.
%
Figure \ref{fig:WMF_a} is the WMF applied on the foot-head velocity component and illustrates perfectly the blood pulse in the ascending and descending aorta.
%axis, we can clearly see that the feature handle perfectly the fluid going on this axis in the aorta.
%
%On the other hand, the fluid part at the peak of the aortic arch is not present at all in this image.
In contrast, $\text{WMF}_\text{u}$ values increases significantly at the top of the aortic arch.
This is a consequence of the absence of velocity in the foot-head direction at this particular location.
%This is due to the fact that there is mainly noises and no fluid dynamics at this point in this axis.
%
The same observation applies to the two figures~\ref{fig:WMF_b} and \ref{fig:WMF_c} 
%(
with, respectively, the left-right and anterior-posterior velocity components.
% axis respectively) for areas where there are no fluid dynamics on the respective axis.
%
%Using the $\text{Min}$ function to merge the three images enables to recover the overall dynamics of the fluid in a single image, as the fluid part has the lower value for WMF.
Therefore, the application of the minimum operator in Eq.~\ref{eq:WMFmin} facilitates the extraction of the most relevant regions from each WMF image.
Due to the Fourier Transform operation, WMF is not a temporal feature.
The proposed feature provides a representation of the voxels which have been exposed to pulsatile flows.

%To go deeper in the analysis of WMF, Figure~\ref{fig:analysis_wmf_c} presents the speed evolution during time and the \seb{FT}
%Fourier Transform 
%of four different points in a \4d.
To further investigate WMF feature, Figure~\ref{fig:analysis_wmf_c} presents the temporal evolution of speed and its Fourier Transform for 4 points of interest located in Figure~\ref{fig:analysis_wmf_a} and~\ref{fig:analysis_wmf_b}.
%
%Points are visible in Figures~\ref{fig:analysis_wmf_a} and \ref{fig:analysis_wmf_b} that shows, respectively, the anatomic and WMF images.
%
Point~\#1 is located at the border of the aorta.
It remains within the aorta and the fluid domain for 17 time frames (illustrated in red on the chart), and subsequently transitions to the exterior domain for the remainder of the time frame.
The %progression of 
speed dynamic reveals a distinct change with a smooth evolution during the period in the fluid domain and unstable velocity measurements outside.
%in its behaviour.
%
%During the period within the fluid domain, the speed evolves smoothly, however, it transitions to a noisy state immediately after exiting this domain.
%
The resulting FT illustrates the higher pulsatility contribution of the $w-$velocity component along with its $\text{WMF}_\text{w}$.
A similar observation can be done for the point \#2 which is positioned in the center of the descending aorta.
% C'est très bien comme analyse, il faut juste que l'on raccourcisse pour le papier 
%The FT of the point shows higher values for the firsts frequencies thanks to the overall behaviour during the in flow part.
%
%Instead, there is no plateau at lower frequencies, which is due to the passage outside the aorta, which has created noise on all three components.
%
%In opposition, the second point is entirely situated within the aorta and the fluid domain.
%
%The behaviour of the fluid is analogous to that observed in 4D flow MRI, with a maximum velocity on the foot-head axis at the initial frames (systolic phase) in the descending aorta due to the contraction of the heart.
%
%The FT demonstrates elevated values for the initial frequencies and a nearly horizontal plateau for the subsequent frequencies.
%
Minor distortions observed in the higher frequencies plateau are attributable to the MRI noise. % present in the MRI.
It results on a small value for WMF which is characteristic of the behaviour of the fluid zone.

%Points~\#3 and \#4
The two last points are %points taken
located fully outside of the fluid domain.
Point \#3 is taken in a high anatomical signal area in the posterior region that results in a low velocity and low noise~\cite{Pelc1991}.
%
%The presence of a high signal in the anatomical image results in low velocity and low noise on the velocity components.
%
%Despite small velocity and noise, chaotic behaviour is observable on the FT with a at low energy.
%
%As we performed a weighted mean on the frequencies, we care more about the behaviour of the FT than the values of it.
%
%The last point
In contrast, Point \#4 is located in the lung, exhibiting a weak anatomical signal and therefore high velocity noise.
%a high presence of noise in the velocity images.
%
In both situations, the FT displays an absence of a high energy peak in low frequencies, resulting in a high WMF.
%FT behaviour and WMF are then closer to the precedent point due to noise presence in the image.

%In the end, we can see that the WMF gives lower values to the fluid domain, then to the points oscillating between fluid and non-fluid domain.
Ultimately, the proposed feature associates low frequencies with the fluid domain and with voxels that exhibit intermittent pulsating flow.
%to the points oscillating between fluid and non-fluid domain.
%
%Finally, the non-fluid domain is found to have the highest values, regardless of whether there is a significant signal in the anatomical image.
Finally, the non-fluid domain has been found to contain the highest WMF values, irrespective of the presence of a significant signal in the anatomical image.

For the sake of simplicity, the term WMF is substituted for the term $\text{WMF}_\text{min}$ in the following sections.
%\begin{figure}[h!]
%    \centering
%    \begin{subfigure}{0.39\linewidth}
%        \centering\includegraphics[width=0.8\textwidth]{mag_point.png}
%        \caption{Anatomic image.}
%    \label{fig:analysis_wmf_a}
%    \end{subfigure}
%    \begin{subfigure}{0.39\linewidth}
%        \centering\includegraphics[width=0.8\textwidth]{WMF_point.png}
%        \caption{WMF image.}
%    \label{fig:analysis_wmf_b}
%    \end{subfigure}\\
%    \begin{subfigure}{\linewidth}
%        \centering\includegraphics[width=\textwidth]{Points_fft.pdf}
%        \caption{Speed and Fourier Transform of each point. Only strictly positive frequencies are showed.}
%    \label{fig:analysis_wmf_c}
%    \end{subfigure}
%    \caption{Speed and Fourier Transform analysis at different point in a \4d.}
%    \label{fig:analysis_wmf}
%\end{figure}

\section{WMF in threshold segmentation task}
\label{sec:manual_seg}

%To demonstrate a use case of WMF, a threshold segmentation task has been performed on multiple \4d with multiple derivations of WMF and the state-of-the-art handcraft feature PC-MRA~\cite{dumoulinPCMRA}.
%
%For the sake of simplicity, the term WMF is substituted for the term $\text{WMF}_\text{min}$ in the following sections.

\subsection{Method}

In order to evaluate the efficacy of WMF in facilitating the segmentation process, a threshold segmentation task is conducted.
In this experiment, a comparative analysis is performed on WMF, PCMRA, and combinations of these features in relation to this task.
%
%For each MRI in the dataset and each feature, the threshold that optimises the IoU between the proposed segmentation and the ground truth is identified. 
For each MRI in the dataset and each combination, the threshold is optimised by seeking the best intersection over union (IoU) between the ground truth and the resulting segmentation.
All features are standardised between 0 and 1, and the threshold is iterated over this range with a step size of $0.02$.
The application of the threshold is determined by the following criteria: a voxel is classified as true if its value is greater than the threshold.
%
%In view of the fact that 
Since WMF assigns a lower value to fluid zones and a higher value to non-fluid zones, the inverse of WMF will be employed here, \emph{i.e.} $1-\WMF$.
In order to establish a comparison between the contributions of WMF and PCMRA to the task, combinations of the two features with anatomy were tested.
%
%The different features and combinations are: $\text{Mag}(t)*(1-\WMF)^8$, $\text{Mag}(t)/\WMF$, $\text{Mag}(t)/\WMF^2$, $\text{PC-MRA}(t)/\WMF$ and $\text{PC-MRA}(t)/\WMF^2$. \simon{on garde ça ?}

\subsubsection{Dataset}
\label{sec:dataset_th}

%In order to assess the task of segmentation, a dataset of 43 4D Flow MRIs is utilised.
The \4d dataset is based on a retrospective study that includes 43 patients who have experienced a cardiovascular event.
The research was carried out following the principles of the Declaration of Helsinki.
%It is made up of two sub-datasets which we will refer as dataset $\mathcal{D}_A$ and $\mathcal{D}_B$, \textit{ie.} complete dataset $\mathcal{D} = \mathcal{D}_A \cup \mathcal{D}_B$.
%
The whole dataset is dedicated to the segmentation task of the aorta and has been labeled by non-expert researchers.
%/ expert / non-medicine ??????????
%
Labelling has been performed using ITK-SNAP (v4.2.2)~\cite{y06itksnap}.
Each MRI is annotated at three distinct time frames: one corresponding to the systolic peak and two frames captured during the diastolic phase, one in the middle of the sequence and the other at the end.
The present dataset includes 26 isotropic (ISO) images in addition to 17 non-isotropic images, with spatial resolutions ranging from $2.5$\,mm to $3$\,mm.
The dataset is separated in 35 1.5\,T acquisitions and 8 ones with a 3T magnetic field.
%\simon{There is 35 1.5T MRIs and 8 3T ones.}
%
Velocity encoding (VENC) changes %fluctuates
within the range of $1.5$ to $6.0\,\text{m}\cdot s^{-1}$ across the dataset.
This dataset is composed of images acquired in different acquisition planes with various contrast enhancement conditions.
The heterogeneity of this dataset is then representative of the clinical condition of acquisition.

% WMF DATASET
%Dataset $\mathcal{D}_A$ is composed of 32 \4d from patients whose presented a cardiovascular event.
%
%Most of MRI present a spatial resolution of $2.5\times 2.5\times 2.5$mm and multiple velocity encoding
%Among the set of MRIs, 29 of them have a spatial resolution of $2.5$mm isotropic (ISO), while the remainder boast a resolution of $3.0$mm ISO or higher.
%
%Velocity encoding (VENC) fluctuates within the range of $1.5$ to $6.0\,\text{mm}\cdot s^{-1}$ across the dataset.
%
%Several types of contrast are present in the set due to the presence or absence of contrast agents or the MRI parameters.
%
%Acquisition plan is also varying in the set and MRIs are not always focused on aorta, some look more about mitral valve or heart.
%
%Due to all these varying parameters, dataset A is more heterogeneous and represents at best the clinical case of \4d.

% PCMRA DATASET
%Dataset $\mathcal{D}_B$ is composed of 11 images from the same type of patients. 
%
%Only two VENC ($2.0$ or $2.5\,\text{mm}\cdot s^{-1}$) are defined and spatial resolution is evolving between $2.2\times 2.2\times 2.0$mm and $2.5$mm ISO in the subset.
%
%All MRI scans were acquired in the same plane (AP\_FH) and gadolinium (contrast agent) was injected into each patient.
%
%This results in a more homogeneous data set that represents ideal situations.

%\sebc{Je suggère de passer en subsubsection et inclure cette partie dans Method. C'est assez logique de donner la méthode d'évaluation.}
\subsubsection{Evaluation and Implementation}

The following evaluation metrics have been selected for the segmentation task: Intersection over Union (IoU), Dice Score, recall and precision.
%To evaluate segmentation, we chose IoU, Dice Score, recall and precision metrics.%, defined as:
%\begin{equation}
%    \text{IoU} = \frac{|\text{Prediction} \cap \text{Label}|}{|\text{Prediction} \cup \text{Label}|}
%\end{equation}
%\begin{equation}
%    \text{Dice Score} = \frac{2*|\text{Prediction} \cap \text{Label}|}{|\text{Prediction}| + |\text{Label}|}
%\end{equation}
%
As it is not a learning task, the evaluation is performed on the whole dataset.
To compare WMF with PC-MRA, the implementation from Bustamante et al. (2018)~\cite{BUSTAMANTE2018128} has been selected, with a gamma of 0.2 that has been shown to improve the visibility of low velocities.
%To compare \seb{WMF with} PC-MRA, we chose the implementation from Bustamante et al. (2018)~\cite{BUSTAMANTE2018128} with a gamma of 0.2 that improves the visibility of low velocities. % j'ai utilisé les memes termes que dans le papier.
%
The peak systolic PC-MRA is, due to the high velocity of this frame, the PC-MRA frame where the fluid domain is the most visible.
As WMF is not a time-dependent feature, the proposed analysis integrates the use of the systolic PC-MRA, denoted PC-MRA (sys), across all the time frames.
Classical PC-MRA computed on each time frame separately is denoted PC-MRA($t$) and Mag($t$) denotes the anatomical image at the time frame $t$.
%Elsewhere, as WMF is not a temporal feature, we chose to compare WMF with the PC-MRA peak systolic frame, refered as PC-MRA (sys).
%
%PC-MRA (sys) is, due to the high velocity of this frame, the PC-MRA frame where the fluid domain is the most visible.
%\sebc{Ajouter une phrase ou deux sur la version PC-MRA (sys) pour montrer que son évaluation est pertinente. C'est aussi le lieu de dire quelle est la méthode (et le papier avec qui on se compare). Vu que ce n'est pas du learning, il faut préciser que l'évaluation a lieu sur toutes données.}

\subsection{Results}

%Results of threshold segmentation are presented in Tables~\ref{tab:seg_th_all} and \ref{tab:seg_th_both}, where results are computed on the dataset $\mathcal{D}$ and on both subsets $\mathcal{D}_A$ and $\mathcal{D}_B$, respectively.
%Results of threshold segmentation are presented in Table~\ref{tab:seg_th_all} for the several feature combinations.
As demonstrated in Table~\ref{tab:seg_th_all}, most of the WMF-derived features manifest better performance on the different metrics than PC-MRA only thresholding.
Out of all the possible combinations, WMF is actually the one that exhibits the best performance.
%Overall, WMF has better performances than other methods on the \seb{whole dataset}
%$\mathcal{D}$ dataset.
% 
Indeed, IoU and Dice Score
%increase
are higher by $0.025$ and $0.034$, respectively, compared to the second best method (7th).
%
%Compared to the state-of-the-art method $\text{PC-MRA}(t)$, the improvement in performance is a factor of 2 for the IoU and 1.9 for the Dice Score.
Compared to the state-of-the-art $\text{PC-MRA}(t)$ method, the WMF-based solution achieves improvements in IoU and Dice Score by factors of 2 and 1.9, respectively.
The recall is more or less uniform across all features, ranging between 0.555 and 0.599 on the entire dataset. 
% $\mathcal{D}$.
%
%% None of the features show any improvement in the proportion of areas of interest found. \sebc{<-Je ne sais pas si cette phrase est nécessaire ici.}
%
%The precision shows that WMF recovers the most important part of the segmentations with an improvement of $0.041$ compared to the second best solution (7th).
%The precision 
Precision metric indicates that WMF recovers the highest proportion of segmentation in relation to the area identified with an improvement of $0.041$ compared to the second best solution (7th).
%
%The results are of the same order for our feature on dataset $\mathcal{D}_A$, which presents more heterogeneous data.
Finally, the systolic PC-MRA applied to the other time frames demonstrates a superior capacity for aorta segmentation compared to the $\text{PC-MRA} (t)$ one.
Although WMF and the systolic PC-MRA are temporally fixed features, WMF has the ability to better represent areas of blood flow.
%For $\mathcal{D}_B$, PC-MRA (sys.) and $\text{PC-MRA}(t)/\WMF^2$ show the best results in IoU and Dice Score.
%
%This is because the contrast in anatomical images is always good in this dataset that has a strong influence on the feature, ie equation.
%
%The PC-MRA (sys.) feature demonstrates a superior capacity for aorta segmentation when compared to the $\text{PC-MRA} (t)$ one.
%
%This is attributable to the fact that the velocity of flow is at its zenith during the systolic peak.
%
%Consequently, the delineation of fluid areas on the feature is enhanced, ie TODO EQUATION, despite the anatomy having moved between peak systolic and another time frame in the MRI.
\begin{table}[bt]
    \centering
    \caption{Results obtained for threshold segmentation on each feature for the complete dataset. \textbf{Best} and \underline{second best} results bold and underlined respectively.}
    \label{tab:seg_th_all}
    \begin{tabular}{|cc||c|c|c|c|}\hline
        \# & Feature & IoU & Dice & Recall & Precision \\\hline
        1 & $\text{PC-MRA}(t)$ & 0.129 & 0.215 & \underline{0.596} & 0.151 \\\hline
        2 & $\text{PC-MRA}$ (sys) & 0.205 & 0.324 & 0.586 & 0.250 \\\hline
        3 & WMF & \textbf{0.262} & \textbf{0.408} & 0.569 & \textbf{0.331} \\\hline
        3 & $\text{Mag}(t)*(1-\WMF)^8$ & 0.225 & 0.358 & 0.564 & 0.271 \\\hline
        4 & $\text{Mag}(t)/\WMF$ & 0.180 & 0.293 & 0.584 & 0.214 \\\hline
        5 & $\text{Mag}(t)/\WMF^2$ & 0.224 & 0.356 & 0.555 & 0.275 \\\hline
        6 & $\text{PC-MRA}(t)/\WMF$ & 0.202 & 0.324 & \textbf{0.599} & 0.241 \\\hline
        7 & $\text{PC-MRA}(t)/\WMF^2$ & \underline{0.237} & \underline{0.374} & 0.584 & \underline{0.290} \\\hline
    \end{tabular}
\end{table}

\section{WMF in segmentation task using deep learning}
\label{sec:deep_seg}

%The following section analyses the usefulness of the feature as input to a neural network for a deep learning segmentation task.

\subsection{Method}

%To analyse WMF's contribution to the deep learning segmentation task, we chose to use the classical U-Net architecture~\cite{ronnebergerUNetConvolutionalNetworks2015} in a 3D configuration.
In order to analyse the contribution of WMF to the deep learning segmentation task, the classical U-Net architecture~\cite{ronnebergerUNetConvolutionalNetworks2015} is used in a 3D configuration, with the time dimension being ignored.
%
%A 3D U-Net is used because the MRI time frames are processed independently, as 3D volumes.
%
The contracting path is constituted by a series of convolution blocks, with each block comprising two convolutions with a kernel size of $3\times 3\times 3$.
Each convolution is followed by a ReLU activation function.
Between each block, a 3D max pooling operation is performed to reduce the resolution size of features with a factor of 2.
The expansive path is made of the same blocks than the contracting one.
Upsampling features between each block is performed thanks to transposed convolution ("Up-Convolution") using a kernel size of $2\times 2\times 2$.
A final $1\times 1\times 1$ convolution and Sigmoid operation are performed to classify the voxels.
U-Net contracting path is made of 4 convolution blocks with, respectively, $[64, 128, 256, 512]$ channels.

%To train the network, a combined loss of Binary Cross Entropy and Dice losses is used, defined as:
%\begin{equation}
%    \mathcal{L} = \mathcal{L}_{\text{BCE}} + \mathcal{L}_{\text{Dice}}
%\end{equation}
% Je retire l'equation car elle n'est pas utilisé ailleurs dans le document. On va gagner un peu de place.
The training is performed based on the sum of the Binary Cross Entropy and Dice losses.
%
%Network is trained using different type of images/features combination as input.
The network is trained using a combination of different types of image/feature as input.
%
%Images and features \seb{combined} %used
%for this task are:
The images and features that have been combined for this task are as follows: Magnitude image (anatomical image, Mag), velocity norm ($\mathcal{V}$), WMF, PC-MRA($t$) and PC-MRA(sys) at the systolic frame.
The input images and features are then concatenated before being processed by the network.

\subsubsection{Dataset}

The same dataset has been used for this task as the threshold segmentation, as referenced in Section~\ref{sec:dataset_th}.
The validation and test sets are each
%both 
composed of 6 MRIs. %, %3 from the $\mathcal{D}_A$ dataset and 3 from the $\mathcal{D}_B$ one to be balanced between the two subsets.
Training set is made of 30 MRIs. % from both subsets.
One of the acquisitions has been removed from the dataset due to incompatibility with the input network dimensions.
%One MRI set from $\mathcal{D}_b$ set is not used as the resolution is too low to be passed through the network.
%
The selection of images to be incorporated within the test and validation sets was undertaken manually.
These sets have been chosen to represent the variety of data (contrast, spike artifacts, acquired planes) by using a stratified sampling method.
%We perform stratified sampling to made a validation and a test set is representative of all type of data within $\mathcal{D}$ set.
%
%We explicitly chose MRIs with low- or high-contrast (due to the presence or absence of gadolinium), spike artifacts or different acquisition planes.
%
For the training and validation sets, 10 patches with a size of $48\times 48 \times 24$ are extracted from each labelled time frame.
As 3 time frames are labelled per MRI, the training set is composed of 900~patches and 180 for validation.
The test set is made of the selected labelled \4d volumes which represents 18 3D images.
%each labelled \4d time frames, it represents 18 3D images.

\subsubsection{Evaluation and Implementation}

To evaluate the network, Dice score and IoU are used to estimate the overlap between predicted aorta segmentation and label segmentation. 
Average and median along the dataset of each metric is given. % TODO: ajouter qqchose ?

%we
The experiment is implemented using Python (v$3.12.6$) and PyTorch (v$2.4.1$).
The network is trained on a Nvidia A100 80~GB and CUDA (v$12.1$).
The SGD optimizer was used, with  a momentum of 0.9 and a weight decay of $10^{-4}$.
The initial learning rate is set at 0.05, and it undergoes a decrease during the training process as follows: $lr = lr_\text{initial} * (1 - epoch_\text{current} / epoch_\text{max})^{0.9}$.
%Initial learning is set at 0.05 and it is decreasing along the training process as:
%\begin{equation}
%    lr = lr_\text{initial} * (1 - epoch_{current} / epoch_{max})^{0.9}
%\end{equation}

%Multiple data augmentation techniques are used with a probability of 0.5, as: random noise, contrast, rotation and elastic transform.
A set of data augmentation techniques is applied with a probability of 0.5, including random noise, contrast, rotation, and elastic transformation.
Random contrast augmentation employs either a random gamma or a random bias field transformation (from the TorchIO library~\cite{perez-garciaTorchIOPythonLibrary2021}).
%For the random contrast augmentation, if applied, there is 50\% of chance to applied a random gamma elsewhere a random bias field (as the MRI artifact).
%
%Random bias field and gamma implementations from the TorchIO library~\cite{perez-garciaTorchIOPythonLibrary2021} are used.

\subsection{Results}

%Table~\ref{tab:seg_dl_all} presents \seb{the performance metrics}
%results 
%for the deep learning segmentation task \seb{with respect of different feature combination}.
% on dataset $\mathcal{D}$.
Table~\ref{tab:seg_dl_all} presents the performance metrics for the deep learning segmentation task with respect to different feature combinations.
\begin{table}[bt]
    \centering
    \caption{Results obtained for deep learning segmentation on the test set. \textbf{Best} and \underline{second best} results bold and underlined respectively.}
    \label{tab:seg_dl_all}
    \begin{tabular}{|ccccc||c|c||c|c|}\hline
        \#&\multicolumn{4}{c||}{Input} & IoU avg. (std) & median & Dice avg. (std) & median \\\hline
        1&$\text{Mag}(t)$ & & & & 0.542 (0.107) & 0.507 & 0.697 (0.090) & 0.672 \\\hline
        2&$\text{Mag}(t)$ & $\mathcal{V}(t)$ & & & 0.513 (0.110) & 0.564 & 0.671 (0.104) & 0.721 \\\hline
        3&$\text{Mag}(t)$ & $\mathcal{V}(t)$&$\WMF$ & & \textbf{0.671} (0.130) & \textbf{0.683} & \underline{0.795} (0.104) & \textbf{0.812} \\\hline
        4&$\text{Mag}(t)$&$\mathcal{V}(t)$ & & $\text{PC-MRA}(t)$ & 0.485 (0.143) & 0.504 & 0.640 (0.138) & 0.670 \\\hline
        5&$\text{Mag}(t)$&$\mathcal{V}(t)$&$\WMF$&$\text{PC-MRA}(t)$ & \underline{0.665} (0.086) & 0.647 & \textbf{0.796} (0.061) & 0.786 \\\hline
        6&$\text{Mag}(t)$&$\mathcal{V}(t)$ & & $\text{PC-MRA (sys)}$ & 0.555 (0.269) & \underline{0.674} & 0.664 (0.288) & \underline{0.805} \\\hline
        7&$\text{Mag}(t)$&$\mathcal{V}(t)$&$\WMF$&$\text{PC-MRA (sys)}$ & 0.625 (0.157) & 0.659 & 0.756 (0.134) & 0.795 \\\hline
    \end{tabular}
\end{table}

The network achieves better results when it uses the WMF feature as input (as shown by the training \#3, 5 and 7). 
%
%WMF improves \4d segmentation, the feature brings information about the fluid domain and helps network to classify correctly voxels.
%
The combination using the peak systolic PC-MRA as input demonstrates optimal outcomes among the non-WMF trainings, attaining the second-best median for both metrics (0.009 and 0.007, IoU and Dice Score, respectively, behind \#3).
%The network utilising PC-MRA, at the systolic peak frame, as input demonstrated optimal outcomes among the non-WMF trainings, attaining the second-best median for both metrics (0.009 and 0.007, IoU and Dice Score respectively, behind \#3).
%
In a low-contrast contexts, segmentation is severely impacted when relying solely on PC-MRA.
%As PC-MRA provide less information while the contrast in the anatomical image is low, it results in very bad segmentation for this type of images.
%
%The gap between the metrics average and median for solution \#\seb{6}
%7 
%demonstrates this, with a difference of 0.119 and 0.141 for the IoU and the Dice Score, respectively.
This is evidenced by the discrepancy between the average and median metrics for training \#6, with a difference of 0.119 and 0.141 for the IoU and the Dice Score, respectively.
%
%This indicates that the network performs well overall with this feature, yet it experiences a complete collapse on certain images characterised by low contrast.
%
Compared to PC-MRA, WMF shows a minimal disparity between the mean and the median (0.012 and 0.017 for the IoU and Dice Score, respectively), which indicates a certain contrast robustness.
%In comparison with PC-MRA, WMF is characterised by the absence of this aspect, exhibiting a minimal disparity between the mean and the median (0.012 and 0.017 for the IoU and Dice Score, respectively).
%
This is attributable to the absence of anatomy in the feature calculation, a factor that renders it significantly more robust to the particularities of the image.
This assertion is corroborated by the observation that the presence of WMF in solution \#7 has resulted in a substantial reduction in the discrepancy between the means and the medians compared to the combination \#6.

\section{Discussion}
\label{sec:discussion}

% on a pu observer que WMF est une feature importante pour la tâche de segmentation par seuil et par apprentissage profond
The threshold and deep learning segmentation experiments showed the importance of WMF feature for the segmentation task.
The information provided by the hull of pulsatile velocity voxels helps the segmentation task and gives information about the fluid domain over time.
% les résultats proposées par les solutions utilisant la feature sont globalement les meilleurs et ne sont pas impactés par la présence ou non d'un fort contraste (agent de contraste) dans l'anatomie, là où PC-MRA est sensible
%Overall, solutions using WMF as input feature give better results and are not impacted by the contrast in the anatomical image, unlike PC-MRA.
% à vérifier avant: entrainement converge plus rapidement avec wmf (c'est globalement le cas comme pcmra sys et pas (t))
%Moreover, we observed, during training process, that network converge faster to the optimal solution while using WMF or PC-MRA (sys) as input.
Furthermore, during the training process it emerged that the network converged more rapidly toward the optimal solution when using WMF or systolic PC-MRA as input.
% Comparaison entre Seg systole ou diastole et impact de la feature (pas trop observé sur la segmentation deep, l'anatomie doit rattraper le tout)
%
%The threshold segmentation task highlights that PC-MRA($t$) is highly influenced by the speed in the image, we observed that Dice Score decrease by a factor of 2.5 between the peak systolic and the end diastolic frame (0.327 and 0.129, respectively).

PC-MRA is a function of both the anatomy and velocity components.
In contrast, WMF is only a function of the velocity itself.
Furthermore, the distribution of fluid domain velocity noise is dependent on anatomical SNR~\cite{Levilly2020}.
The anatomical and velocity SNR should have an effect on both PC-MRA and WMF.
However, an observation of the deep learning test dataset (comprising only six volumes) revealed a reduced sensitivity to the SNR for the proposed feature WMF.
This observation requires confirmation using a larger dataset.

%The Signal to Noise Ratio (SNR) of the fluid domain has an influence on WMF.
%
%In fact, a low SNR in the fluid domain results in a higher WMF value in this area (as the outflow domain) due to the higher noise distribution over time.
%
%Overall, solutions using WMF as input feature give better results and are not directly impacted by the anatomical image, unlike PC-MRA.
%
%In fact, if the SNR is low in the fluid domain, WMF will have a higher value (as the outflow domain) for this area due to the noise distribution over time.
%
%\seb{Besides, an MRI acquisition combining a low SNR and a high VENC ($6\text{m}\cdot\text{s}^{-1}$) led to degraded performances with PC-MRA and WMF in the threshold segmentation task.}
%A low SNR has influence on PC-MRA as well, we observed it during the threshold segmentation task where a MRI with a high VENC ($6\text{m}\cdot\text{s}^{-1}$), and so on a low SNR.
%
%\seb{In that set,}
%If we consider a specific MRI in this condition (a low SNR and a high VENC, $6\text{m}\cdot\text{s}^{-1}$), performance decreases but 
%WMF gets a higher Dice Score than PC-MRA (sys) ($0.224$ and $0.102$ respectively, over the three time frames). 

%For instance, the results of the threshold segmentation task demonstrate that PC-MRA($t$) is significantly affected by image speed during the diastolic period (\emph{i.e.} with a smaller velocity SNR).
Threshold segmentation results indicate that PC-MRA($t$) is notably affected by image speed during diastole, corresponding to reduced velocity SNR.
%
%A decrease in the Dice Score was observed by a factor of 2.5 between the peak systolic and the end diastolic frame (0.327 and 0.129, respectively).
%\simon{The PC-MRA Dice Score decreases by a factor of 2.5 between the peak systolic and diastolic frames (0.327 and 0.129, respectively), while WMF remains stable (0.409 and 0.407).}
A 2.5-fold decrease in the PC-MRA Dice score is observed between peak systole and diastole (0.327 vs. 0.129), whereas WMF values remain stable (0.409 vs. 0.407).
%
%WMF being function of several timeframe velocity values and designed to reveal the pulsatility pattern, it is then less sensitive to velocity SNR.
WMF, as it incorporates velocity information over several timeframes to characterize pulsatility, exhibits reduced sensitivity to velocity SNR.
%On the other hand, WMF remains unaffected by this phenomenon due to the feature definition (0.409 and 0.407 for the same frames).

%A pardon, tu fais un focus une IRM.
%\sebc{<- pas pour moi, je ne sais pas de quel contexte tu parles. Il me semblait que l'on avait évoqué l'idée de comparer les performances entre systole et diastole pour WMF (et peut être PC-MRA (sys)).}

% des résultats interessants sont données par les solutions combinant PCMRA et WMF, ce qui indique qu'il y a peut etre quelque chose à creuser par là, que les features ne sont pas forcement à mettre en opposition
During the experiments, combinations of PC-MRA and WMF has been tested and provide interesting
%good 
results (\#7 for threshold segmentation and \#5 for deep learning one).
PC-MRA and WMF are not opposed to each other, each provides different information.
PC-MRA is a handcraft feature
%technique 
that highlights the high anatomy and velocity values
%fluid zones 
in any \4d frame.
WMF, on the other hand, represents the hull of all pulsatile velocity voxels.
% de plus, la feature permet de représenter pulsalité globale de l'image, qui est une facon nouvelle de représenter l'irm de flux 4d -> possibilité d'usage pour l'analyse clinique
It is a new way to represent \4d images.
Besides, it could be used as biomarker in clinical analysis
%can have a clinical usage 
to have a better understanding of the fluid domain and the pulsatility. % Surement refaire la phrase, je n'ai pas trouvé de bon rewrite avec Deepl

\section{Conclusion}
\label{sec:conclusion}

% Dans cette étude, nous proposons la nouvelle feature WMF qui permet la représenté la hull truc là
In this study, a novel handcraft feature, called Weighted Mean Frequencies (WMF), is presented for \4d.
WMF has been introduced with the objective of assisting the task in differentiating between pulsatile and non-pulsatile fluid regions.
Classical PC-MRA and WMF provide complementary information.
WMF showed the capacity to improve results of \4d segmentation using a threshold or a deep learning approach.
%
%The feature helps the network to find the fluid domain and converge faster during training phase.
This feature helps the network find fluid domains and converge faster during the training phase.
%
%Despite the feature advantages, WMF is sensitive to the SNR, and more specifically the noise distribution in the fluid domain.
%
%\sebl{Moreover, PC-MRA and WMF provide complementary information, and their combination should be the subject of further study.}
%
%But unlike PC-MRA, the low contrast MRIs can be handle easily as the anatomical image is not use to compute the feature.
%JE VOULAIS PAS FINIR CE PARAGRAPHE SUR UNE NOTE NEGATIVE DONC J'AI AJOUTE CA

% L'utilisation de WMF pour la segmentation devrait être investiguer plus en profondeur avec des architectures plus récente like approches using transformer or mamba block
Moreover, deep learning segmentation using WMF as input should be deeper investigated by using more recent approaches as Transformer~\cite{9706678,10526382} or Mamba~\cite{U-Mamba,liao2024lightmunet}.
% utilisation de WMF dans un cadre clinique pour trouver les zones fluides dans les irm de mauvaises qualités ou en tant que biomarker, pour quantifier la pulsatilité du sang.
WMF seems to exhibits a low sensitivity to the velocity SNR.
Due to the size of the test dataset, an investigation on a larger clinical dataset should be done with more patients and MRI manufacturers.
%would be done.
%
Finally, WMF could be used as a pulsatility biomarker or a tool to detect fluid domain within low quality \4d.

%
% ---- Bibliography ----
%
% BibTeX users should specify bibliography style 'splncs04'.
% References will then be sorted and formatted in the correct style.
%
\bibliographystyle{splncs04}
\bibliography{refs}
\end{document}